# TIME SERIES ANALYSIS ON STOCK MARKET FOR TEXT MINING CORRELATION OF ECONOMY NEWS

**Sadi Evren SEKER**
Computer Engineering Dept.
Istanbul University
Asst. Prof. Dr.
academic@sadievrenseker.com

**Cihan MERT**
Electrical Engineering Dept.
The University of Texas at Dallas
Researcher, PhD.
cxm121131@utdallas.edu

**Khaled Al-NAAMI**
Computer Science Dept.
The University of Texas at Dallas
MSc.
kma041000@utdallas.edu

**Nuri OZALP**
Turkish National Science Foundation
Researcher
nuri.ozalp@tubitak.gov.tr

**Ugur AYAN**
Turkish National Science Foundation
Senior Researcher, Dr.
ugur.ayan@tubitak.gov.tr

─Abstract─

This paper proposes an information retrieval method for the economy news. The effect of economy news, are researched in the word level and stock market values are considered as the ground proof.

The correlation between stock market prices and economy news is an already addressed problem for most of the countries. The most well-known approach is applying the text mining approaches to the news and some time series analysis tech-niques over stock market closing values in order to apply classification or clustering algorithms over the features extracted. This study goes further and tries to ask the question what are the available time series analysis techniques for the stock market closing values and which one is the most suitable? In this study, the news and their dates are collected into a database and text mining is applied over the news, the text mining part has been kept simple with only term frequency – inverse document frequency method. For the time series analysis part, we have studied 10 different methods such as random walk, moving average, acceleration, Bollinger band, price rate of change, periodic average, difference, momentum or relative strength index and their variation. In this study we have also explained these techniques in a comparative way and we have applied the methods over Turkish Stock Market closing values for more than a 2 year period. On the other hand, we have applied the term frequency – inverse document frequency method on the economy news of one of the high-circulating newspapers in Turkey.

**Acknowledgement**

This study is supported by Istanbul University, research projects department under project number YADOP-27254

## 1. INTRODUCTION

This study is built on one of the high-circulating newspapers in Turkey, which have special pages for economy news. We have collected only economy news, which are separate from other news like sports or magazine, etc. The properties of the dataset will be explained in the experiments section. We have processed the news text via the text mining approach called term frequency - inverse document frequency (TF-IDF), which will be explained in the methodology section. On the other hand, we have processed the stock market closing values by using several signal processing approaches such as random walk (RW), relative strength index (RSI), momentum, moving average (MA), difference (DIF), periodic average (PA), Bollinger band (BB), periodic rate change (PRC) and their variations with some optimizations on the parameters. All these methods will be explained in the signal processing chapter, which is under the background chapter. Finally we have investigated the correlation between the features extracted from text mining and signal processing to compare the effect of signal processing outputs into the economy news. During this correlation study, we have implemented k-nearest neighborhood (KNN) and support vector machine (SVM) algorithms, which are discussed in the section of classification. Also this paper holds the implementation details and the methodology of evaluation over both the signal processing and classification results which are held in the evaluation section.

## 2. PROBLEM STATEMENT

This study is the first time to address the correlation effect of signal processing methods on stock market values with the economy news.

**Figure 1. Overview of Study**

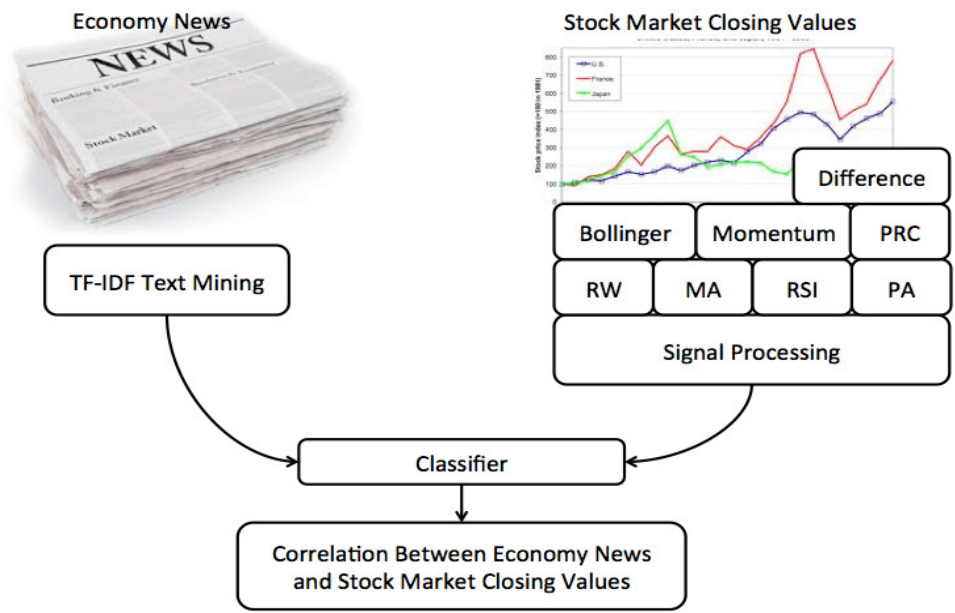

The correlation between news and the stock market is one of the indicators of the speculative markets (Nikfarjam, Emadzadeh, & Muthaiyah, 2010).

One of the difficulties in this study is dealing with natural language data source, which requires a feature extraction. The other difficulty is dealing with a stock market value, which is considered as a signal. On the other hand, the big data we are dealing with is also problematic. The dataset holds 131,248 distinct words and when the feature vector of each economy news item is collected, the total size of the feature vector is over 2.5 Gbyte, which is beyond the computation capacity of a single computer with these classification algorithms.

## 3. RELATED WORK

While there is a great amount of published articles about machine learning, time series, other classification and regression methods in prediction of stock market prices, the numbers of articles involving the application of finding a correlation between stock market and daily news using text mining techniques are poor. News, especially economy news-based stock market prediction, can be considered as a text classification/mining task (Seker & Diri). The main aim is to predict some aspects of the stock market such as price or volatility based on the news content or derived text features. Based on forecasting goal, a set of final classes are defined, such as "Up" (increase in price), "Down" (decrease in price), "Balance" (no change in price) and etc. All proposed algorithms are supposed to classify the incoming news into these classes.

One of the first works on forecasting stock market uses hints (statistics, ratios and interpretations of trend-charting techniques) listed by domain experts to predict weekly UP's and DOWN's of the stock market (Braun & Chandler, 1987). Generally, published related works are built around a central learning algorithm (mostly a classifier) for predicting the sentiment (price direction) of news articles. Wutrich, Cho, Leung, Permunetilleke, Sankaran, and Zhang (1998) try to predict stock market indices using information contained in articles published in The Wall Street Journal (www.wsj.com), Financial Times (www.ft.com), Reuters (www.investools.com), Dow Jones (www.asianupdate. com), and Bloomberg (www.bloomberg.com) based on k-NN learning algorithm and the Euclidean similarity measure.

Similarly, Mittermayer (2004) tries to predict price trends (incline, decline or flat) immediately after press release publications. "Good news" articles are categorized as inclines if the stock price relevant to the given article has increased with a peek of at least +3% from its original value at the publication time at some point during the 60 minutes that follows. The average price level in this period has to be at least 1% above the price at publication. For "Bad news" the opposite is true, the movements have to be in the negative direction instead. The rest of the articles in between are classified as "No movers".

Schumaker and Chen (2009) studied the effect of financial news articles on three different derived text mining features: Bag of Words, Noun Phrases, and Named Entities and their ability to guess discrete number stock prices after an article re-lease, which is a regression problem and not a categorization problem. In this study, Named Entities features scheme gives better results than the Bag of Words on 1-class SVM algorithm.

Mittermayer and Knolmayer (2006) proposed an automated text categorization system using a hand-made thesaurus to forecast intraday stock price trends from information contained in press releases. This system includes three engines: the document preprocessing Engine automatically preprocesses incoming press releases; the categorization engine sorts the press releases into different categories and; finally, the trading engine triggers trading recommendations for the corresponding security. System creates the feature list with the Bag of Words method using either trivial functions like Collection Term Frequency (CTF), Inverse Document Frequency (IDF), and CTF x IDF or ATC-specific functions like Chi-squared (CHI), Information Gain (IG), and Odd's Ratio (OR).

Tan (n.d.) developed a methodology which uses some time series segmentation techniques to find stock price trends by reducing the noise in the price curve. Two different methods of labeling the news articles were introduced. Method one is "observed time lag", which has a time lag between publishing time and when the public absorbs and acts on it. The second method is built on the efficient market theory, which states that there is no time lag between publishing time and market correction for this piece of information.

Schumaker & Chen (2009) proposed another learning system named The Arizona Financial Text System that uses financial news articles and stock quotes as its input to predict future stock price movements. This system uses proper nouns that occur three or more times to be included in the feature set. It distinguishes itself by using a support vector regression (SVR) algorithm instead of a classification in order to predict a future price value when given an article.

Tan (n.d.) describes a classifier based on labeling by trading volume or by returns that can only predict whether a news article will initiate a high number of trades or not; it cannot predict the direction of the price movement. The division of the volume labels is controlled by whether the given volume is above or below the average trading volume. Labeling by returns looks for abnormal return amounts and labels news articles from that.

## 4. BACKGROUND

We have implemented TF-IDF and signal processing methods as already explained in the introduction; this section will discuss these methods in detail. Also one of the difficulties is the number of words we are dealing with. We have implemented the information gain calculation for eliminating some of the features.

**4.1. Term Frequency – Inverse Document Frequency**

TF-IDF is one of the text mining methods used for feature extraction from natural language data sources (Seker et. al., 2013) (Mittermayer, 2004), (Schumaker & Chen, 2009) (Halgamuge, Zhai, & Hsu, 2007) (Fung, Yu, & Lam, 2002).

For the TF-IDF calculation is given in equation (1).

$$tfidf(t, d, D) = tf(t, d) \times idf(t, D) \qquad (1)$$

Where t is the selected term, d is the selected document and D is all documents in the corpus. Also TF-IDF calculation in above formula is built over term frequency (TF) and inverse document frequency (IDF), which can be rewritten as in equation (2).

$$tf(t, d) = \frac{f(t, d)}{\max\{f(w, d) : w \in d\}} \qquad (2)$$

where f is the frequency function and w is the word with maximum occurrence. Also the formulation of IDF is given in equation (3).

$$id(t, D) = \frac{\log(|D|)}{(|\{d \in D : t \in d\}|)} \qquad (3)$$

where |D| indicates the cardinality of D, which is the total number of documents in the corpus.

**4.2. Financial Time Series Analysis**

The stock market closing values are collected from the web page of the Istanbul Stock Market which is available for public download and usage.

**Figure 2. Istanbul Stock Market closing values for 2 years**

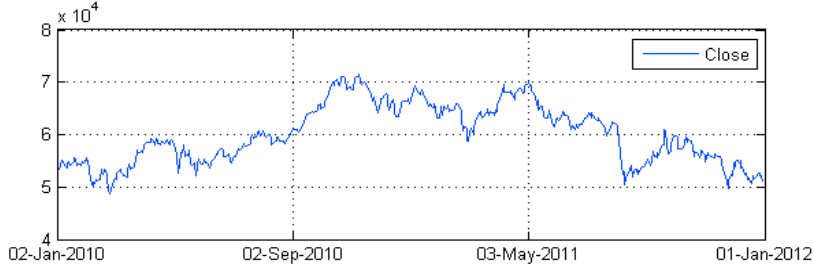

As demonstrated in Figure 2, the values are in local currency which is Turkish Lira and we do not care about. This section focuses on the time series analysis methods. We will try to briefly explain the methods and their variations.

### 4.2.1. Random Walk

Instead we use a relative feature extraction method and get the difference between two consecutive closing values like for any $C_i \in C$ , where C is the set of closing values, we collect pairs of <$C_t$ , $C_{t-1}$> for each news where t is the publication date of the news.

The collected features then subtract from each other to see if there is an increase or decrease in the closing value of the stock market as in equation (4).

$$\text{Feature} = C_t - C_{t-1} \qquad (4)$$

If the feature value of the closing values of the stock market is positive (+), we consider it as 1; if the value is negative (-) we consider it as -1. Also there are some dates which do not hold the closing values such as weekends. For those dates we consider the class label of the news as 0.

This approach can be considered as random walk in the literature. The formulation of random walk is given as in equation (5) (Keikha, Carman, & Crestani, 2009).

$$s_n = \sum_{j=1}^{n} Z_j \qquad (5)$$

where Zj is the random event and the initial value of walk starts by 0 where S0 = 0. The n in the equation (5) determines the length of the walk and in our study the length of walk is limited with only 2 for each cases. Another variation of random walk is extending the walk length. We have tried both methods as separate signal processing approaches in our study.

### 4.2.2. Moving Average

Moving average is a financial analysis model which can be applied on the time series like the stock market values. There are several variations of the method like simple moving average (SMA), cumulative moving average (CMA), weighted moving average (WMA), or exponential moving average (EMA). During this study we have used SMA for the RSI calculation, which will be explained in the next subsection and moving average convergence / divergence (MACD) as a separate method (Karaman & Altiok, 2004).

The simplest form of moving average is calculating the mean of the given time period as demonstrated on equation (6).

$$SMA_{t,n} = \sum_{i=t-n}^{t} \frac{P_i}{n} \qquad (6)$$

Where t is the current time value calculated and n is the window size on the time series and Pi is the price value on time i.

Also the SMA calculation can be followed in an interative manner after the initial SMA value is calculated. WMA on the other hand defines some weights to the price values (Pi) depending on their distance to the current time (t).

$$WMA_{t,n} = \sum_{i=t-n}^{t} \frac{(n+i-t)P_i}{n} \qquad (7)$$

The only difference between the WMA in equation (7) and SMA in equation (6) is the coefficients of the price values $P_i$ where the value gets higher when the date of price gets close to the current time t.

Another variation of moving average is the exponential moving average (EMA) and the coefficients increase exponentially when the price time gets closer in the time series to the current time t.

$$EMA_{t,n} = \sum_{i=t-n}^{t} \frac{(1-\alpha)^{n+i-t} P_i}{n} \qquad (8)$$

In the equation (8) the only difference is the coefficients are getting exponential and the order of the coefficients is getting higher when the calculated price value gets closer to the calculation point on the time series. Also the value of α is the indicator of the strength of the coefficients, where higher α values indicate less importance of the older prices. Originally EMA doesn't deal with the number of prices handled in the window size, instead the α value determines the number of significant coefficients, but because of the implementation limitations, we have limited the number of window size by a constant variable.

Moving average convergence / divergence (MACD) is another method of analysis built on MA and deals with two different time length on the series, which can be named as short and long periods.

The short and long terms are built on the n parameter of EMA. For example in our application we have short as 12 days and long as 26 days and the output is shown in Figure 3.

**Figure 3. MACD values of Istanbul stock market closing values**

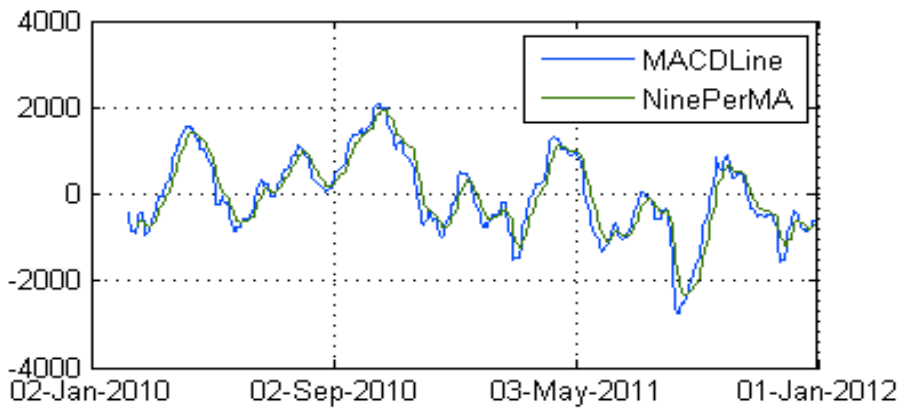

In Figure 3, the NinePerMA values are the EMA9 which is the exponential moving average for 9 days.

### 4.2.3. Relative Strength Index

Relative strength index (RSI) (Wilder, 1978) is a method built over the exponential moving average (EMA) which is already explained in the "moving average" sub section. For two consecutive closing values, e.g. $C_{now}, C_{previous} \in C$, where C is the time series of the closing values in the stock market, there are three possibilities, which are $C_{now} < C_{previous}$ or $C_{now} > C_{previous}$ or $C_{now} = C_{previous}$, for all these possibilities the RSI calculation can be demonstrated as in the equation (10).

$$RSI = 100 - \frac{100}{1 + RS} \qquad (10)$$

where RS as in equation (11).

$$RS = \frac{EMA_{U,n}}{EMA_{D,n}} \qquad (11)$$

where EMA is the exponential moving average as explained in the moving average subsection and $U$ and $D$ values are as in the equation (12).

$$\begin{cases} U = C_{now} - C_{previous}, D = 0, if\ C_{now} > C_{previous} \\ U = 0, D = C_{previous} - C_{now}, if\ C_{now} < C_{previous} \\ U = 0, D = 0, if\ C_{now} = C_{previous} \end{cases} \qquad (12)$$

The RSI values after processing over the Istanbul stock market closing values are demonstrated on figure 4.

**Figure 4. MACD values of Istanbul stock market closing values**

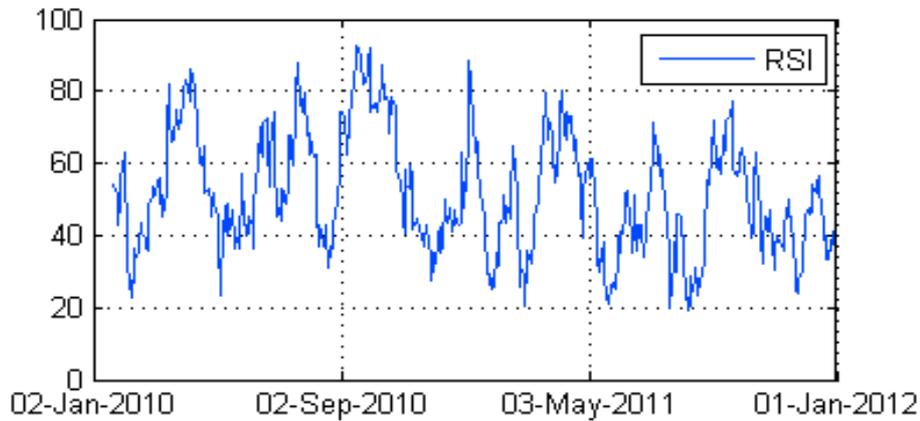

### 4.2.3. Momentum and Rate of Change

Both time series analysis methods are similar to each other. Momentum aims to calculate the change on the *n* day interval and can be calculated as in the equation (13) (AsiaPac Finance , 2013)

$$momentum_n = C_{now} - C_{now-n} \qquad (13)$$

Similarly rate of change can be calculated as the ratio of momentum over the interval as in the equation (14) .

$$rate\ of\ change_n = \frac{momentum_n}{C_{now-n}} \qquad (14)$$

or if the momentum value is substituted the equation (14) can be rewritten as in the equation (15) (AsiaPac Finance , 2013)

$$rate\ of\ change_n = \frac{C_{now} - C_{now-n}}{C_{now-n}} \qquad (15)$$

**Figure 5. Momentum values of Istanbul stock market closing values**

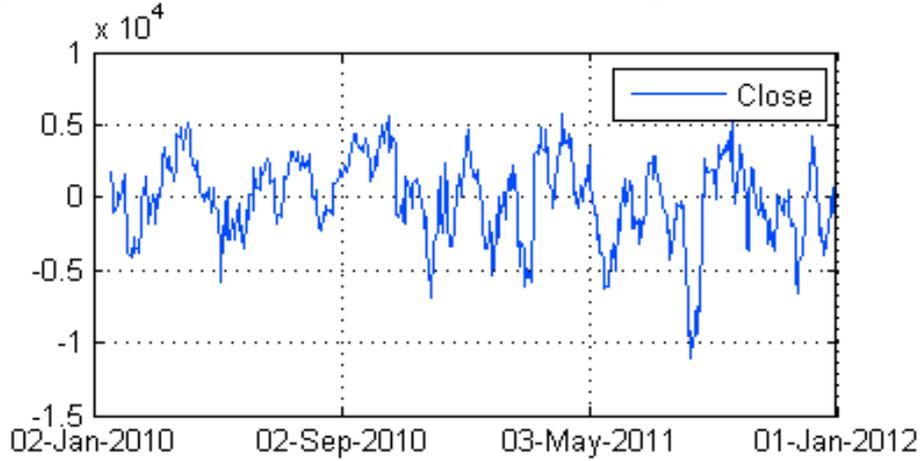

Momentum values are demonstrated in Figure 5 and the ROC values are demonstrated in Figure 6.

**Figure 6. ROC values of Istanbul stock market closing values**

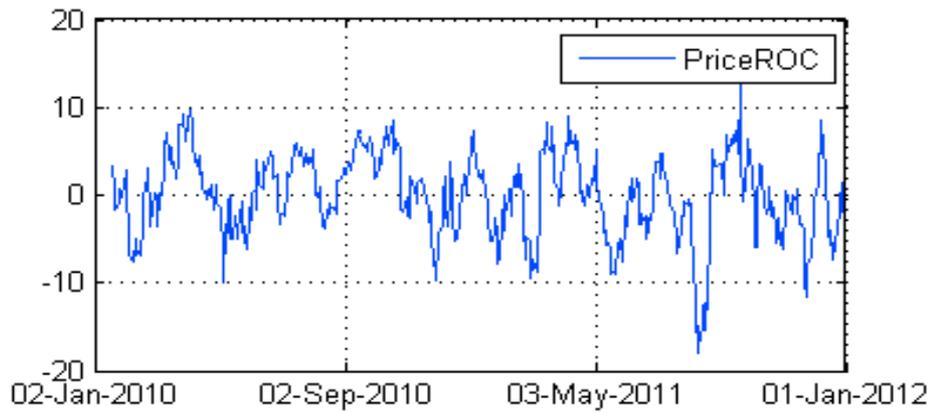

### 4.2.4. Bollinger Band

Bollingber band processing yields a channel of upper, lower and middle bollinger bands. The calculations can be done as in equation (16) (AsiaPac Finance , 2013).

$$BB_{middle,n} = MA_n$$
$$BB_{upper,n} = \sum_{i=t-n}^{t} 1, if\ MA_n > K\sigma_n \qquad (16)$$
$$BB_{lower,n} = \sum_{i=t-n}^{t} 1, if\ MA_n < K\sigma_n$$

The $\sigma_n$ values in the equation (16) indicate the standard deviation for the given time interval *n*. Also the *K* value is a constant which is taken as 20 in our case.

**Figure 7. Middle BB values of Istanbul stock market closing values**

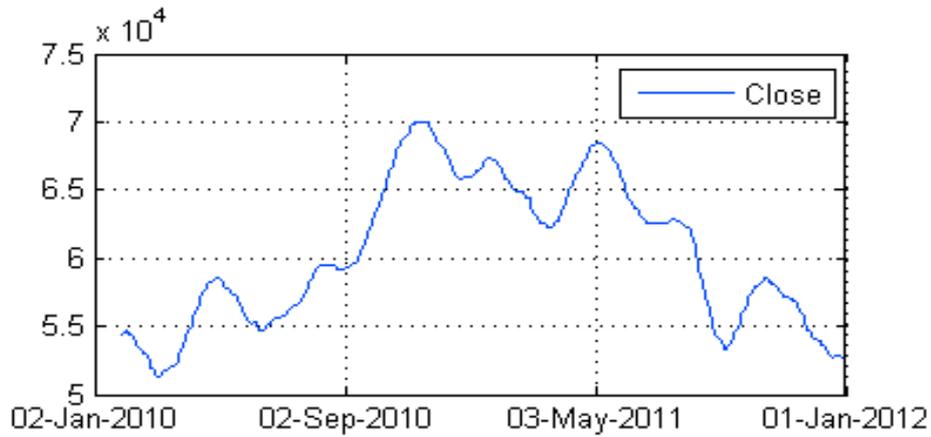

### 4.2.5. Acceleration and Difference

Time series acceleration is another method of analysis. The method can be formalized as in the equation (17) (AsiaPac Finance , 2013).

$$Acc = diff(MACD_n, Histogram) \qquad (17)$$

The approach of acceleration is getting the difference between the moving average convergence / divergence and the histogram of the time series which is the signal itself.

The difference (or convergence / divergence) between two time series (or signals) is considered a subtraction between two series day by day.

The histogram on the other hand is the difference between MACD and the signal itself.

**Figure 8. The output of acceleration over the stock market closing values.**

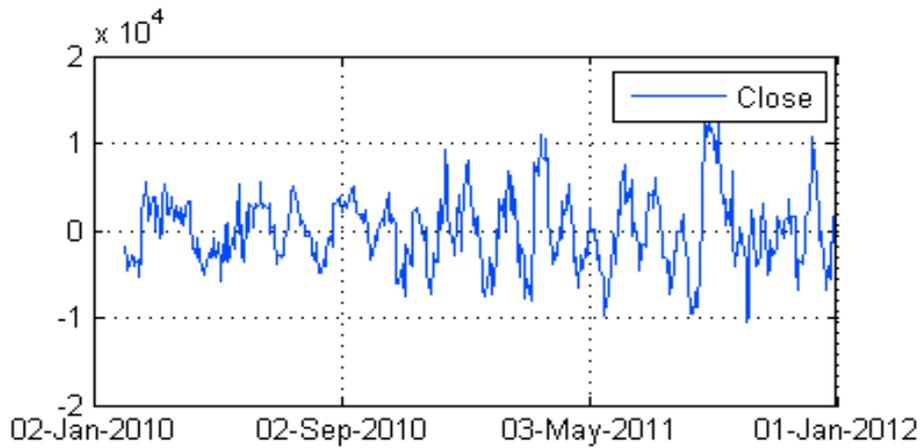

**Figure 9. The time series analysis named "difference", applied over the stock market values.**

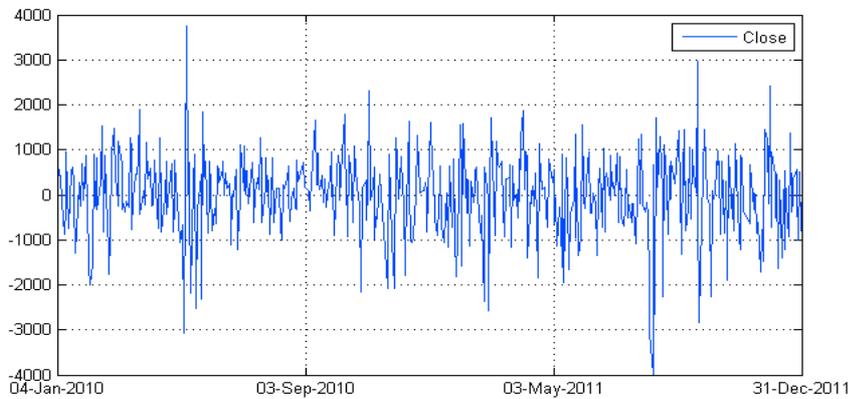

Please note that the acceleration is smoother because of the average calculation taking a part in the formulation.

### 4.3. Information

The information gain of all the terms is calculated and ordered in descending order. Let *Attr* be the set of all attributes and $E_x$ be the set of all training examples, *value(x,a)* with x $\in$ $E_x$ defines the value of a specific example *x* or attribute $a \in$ *Attr, H* , specifies the entropy. The information gain for an attribute $a \in$ *Attr* is defined as in equation (18).

$$IG(Ex,a) = H(Ex) - \sum_{v \in v(a)} \frac{|x \in Ex | v(x,a)|}{|Ex|} H(x \in Ex|v(x,a)) \qquad (18)$$

Also entropy in the information gain calculation can be rewritten as in equation (19).

$$H(X) = \sum_{i=1}^{n} P(x_i) I(x_i) = \sum_{i=1}^{n} P(x_i) \log_b \left(\frac{1}{P(x_i)}\right) = \sum_{i=1}^{n} P(x_i) \log_b (P(x_i)) \qquad (19)$$

### 4.4. K- Nearest Neighborhood (KNN)

The k, c-neighborhood (or k, c(x) in short) of an U-outlier $x$ is the set of $k$ class $c$ instances that are nearest to $x$ (k-nearest class c neighbors of x). The U-outlier is the members that does not fit into any of the classes and considered as an otlier without any classification.

**Figure 10. Visualization of K-NN**

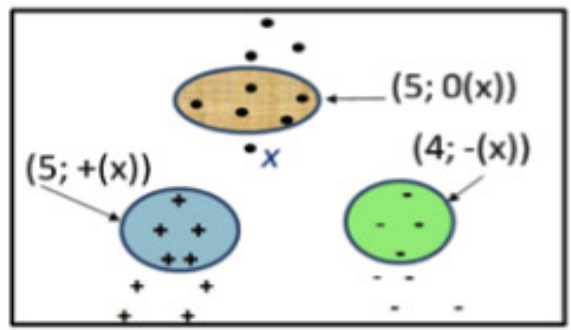

(q; c(x)) **neighborhood** = q-nearest neighbors of x within class c
(5; +(x)) **neighborhood** = 5-nearest neighbors of x within class +

The K-NN (Masud, et al., 2011) is explained in Figure 10. Here $k$ is a user defined parameter. For example, $k$, $c_1(x)$ of an U-outliers $x$ is the k-nearest class $c_1$ neighbors of $x$.

Let $\bar{D}_{Cout,q}(x)$ be the mean distance of a U-outlier $x$ to its k-nearest U-outlier neighbors. Also, let $\bar{D}_{C,q}(x)$ be the mean distance from $x$ to its $k,c(x)$, and let $\bar{D}_{Cmin,q}(x)$ be the minimum among all $\bar{D}_{C,q}(x)$, $c \in$ {Set of existing classes}. In order words, $k$, $c_{min}$ is the nearest existing class neighborhood of $x$. Then k-NSC of $x$ is given in equation (20).

$$k - NSC(x) = \frac{\overline{D}_{C_{min},q}(x) - \overline{D}_{C_{out},q}(x)}{\max\left(\overline{D}_{C_{min},q}(x), \overline{D}_{C_{out},q}(x)\right)} \quad (20)$$

### 4.5. Error Rate Calculation

The error rate of the system is calculated through root mean square error (RMSE). The calculation of RMSE is given in equation (21) (Ocak & Seker, 2011).

$$x_{rmse} = \frac{\sqrt{x_1^2 + x_2^2 + ... + x_n^2}}{n} \quad (21)$$

For this study, above $x$ values are the results achieved from the implementation of the algorithm. The RMSE result of 0 is considered ideal and lower values close to 0 are relatively better.

By the results fetched from the output layer and the calculation of RMSE, the algorithm back propagates to the weight values of the synapses.

Also the results are interpreted by using a second error calculation method RRSE (Root Relative Squared Error) and the calculation is given in equation (22) (Qureshi, Mirza, & Arif, 2006).

$$x_{rrse} = \sqrt{\frac{\sum_{j=1}^{n}(P_{ij} - T_j)}{\sum_{j=1}^{n}(T_j - \overline{T})^2}} \quad (22)$$

Where $P_{ij}$ is the value predicted for the sample case, $T_j$ is the target value for sample case $j$ and $\overline{T}$ is calculated by equation (23).

$$\overline{T} = \frac{1}{n}\sum_{j=1}^{n} T_j \quad (23)$$

The RRSE value ranges from 0 to $\infty$, with 0 corresponding to ideal.

The third error calculation method is RAE (Relative Absolute Error) and the calculation is given in equation (24) (Ocak & Seker 2012).

$$E_i = \frac{\sum_{j=1}^{n} |P_{(ij)} - T_j|}{\sum_{j=1}^{n} |T_j - \overline{T}|} \quad (24)$$

For a perfect fit, the numerator is equal to 0 and $E_i$=0 So, the $E_i$ index ranges from 0 to infinity, with 0 corresponding to the ideal.

Also the success rate of prediction and expectation can be measured as the f-measure method. The f-measure method is built on the Table 1.

**Table 1. f-measure method demonstration**

|  | Predictions |  |  |
|---|---|---|---|
| Expectations |  | Positive | Negative |
|  | True | True Positive | True Negative |
|  | False | False Positive | False Negative |

The calculation of f-measure can be given as in equation (25) depending on the Table 1.

$$F_{measure} = \frac{2TP}{2TP + FN + FP} \quad (25)$$

## 5. EXPERIMENTS

In this study the dataset is in natural language and some preprocessing for the feature extraction from the data source is required. The first approach is applying the TF-IDF for the all terms in the data source. Unfortunately the hardware in the study environment was not qualifying the requirements for the feature extraction of all the terms in data source which is 139,434.

### 5.1. Dataset

We have implemented our approach and Table 2 demonstrates the features of the datasets.

**Table 2. Properties of the Dataset**

|  | News |
|---|---|
| # of News | 9871 |
| Authors | 224 |
| Texts per Author | Mean ($\mu$) : 44.05 Stddev($\sigma$) : 535.52 |
| Average word | ~6.7 |

| length | |
|---|---|

The above dataset is collected from the web site of a high-circulating newspaper in Turkey. The data is collected directly from a database so the noisy parts on the web page like ads, comments, links to other news, etc. are avoided. Another problem is the noise of HTML tags in the database entries for formatting the text of news. The data has preprocessed and all the HTML tags are removed from the news and also all punctuations and stop words are removed in the preprocessing phase.

### 5.2. Feature Extraction

We have implemented a feature extraction algorithm 1 in order to extract two feature vectors.

**Algorithm: Feature Extraction Methods**

1. *Let E be Economy News Corpus,*
2. *Let C be Closings of Stockmarket,*
3. *For each $E_i \in E$*
4.    *For each $Term_j \in E_i$*
5.       *if(count($Term_j$)>30)*
6.          *$T_j \leftarrow$ TF-IDF of $Term_j$*
7.       *$C_i \leftarrow$ closing_value(date($E_i$)) $\in C$*
8. *$IG_{ij} \leftarrow$ Information Gain ($Term_j,E_i$)*
9. *$V_1 \leftarrow$ Top300(sort(IG))*
10.   *$V_2 \leftarrow C$*

The above algorithm demonstrates the extraction of two vectors: one from the economy news corpus and another from the closing values of the stock market. We have limited the number of features to 300 and the Top300 function gets the topmost 300 features from the feature vector.

The $V_2$ feature vector is calculated easily by checking the closing value of the economy news on the date. There are some news items which are published during the time the stock market is closed like on weekends and we have considered these values as a third class besides the increase and decrease classes.

The correlation algorithms run over the two vectors $V_1$ and $V_2$ extracted via the Algorithm 1.

During the execution of algorithm, the execution requires more memory than the available hardware, where we run the algorithms on a intel 7 cpu and 8GByte of RAM. The required memory is calculated in equation (26).

*Memory Requirement = 139,434 words x 9871 news x 6.7 average word length x 2 bytes for each character*

*=~17GByte* (26)

As a solution we have limited the number of words with the highest occurences. The number of occurences on our implementation is 30 and a word is taken into consideration after this number of occurences. The words appearing above this threshold value are 2878 and the memory required is reduced to 700MByte which is easier to handle in the RAM.

The feature vector extraction is about 56 minutes on average for the economy news.

### 5.3. Classification

In the classification step, we have implemented the 10-fold cross validation method, which uses the first 90% as training set and rest 10% as test set. After completing the classification and getting the results, the test set shifts to the next 10% in circular manner until all the instances in the data set gets a part of test set once. The total number of train / test sepration is ten times with a different part is considered as test in each time.

The results of executions can be summarized in Table 3.

**Table 3. Error and Success Rates of Classification Methods**

|  | f-measure Average | RMSE | RAE | Correctly Classified |
|---|---|---|---|---|
| Random Walk | 0.508 | 0.4398 | 0.9924 | 51.58% |
| Bollinger Band | **0.529** | **0.4383** | **0.9883** | **52.92%** |
| Moving Average | 0.507 | 0.4375 | 0.9937 | 52.07% |
| Momentum | 0.504 | 0.4436 | 0.9974 | 50.37% |
| Difference | 0.505 | 0.444 | 1.0009 | 50.45% |
| RSI | 0.501 | 0.4404 | 0.9930 | 50.70% |
| ROC | 0.505 | 0.4422 | 0.9933 | 50.86% |
| *Acceleration* | *0.968* | *0.1543* | *0.9752* | *96.74%* |
| Periodic Average | 0.508 | 0.443 | 0.9960 | 50.77% |
| Random Walk Length=2 | **0.297** | **0.585** | **0.9627** | **37.67%** |

The success rate in Table 3 is the percentage of correctly classified instances. For example, the success rate of Random Walk with length=2 can be considered as the 37% of the instances are correctly classified to predict an increase, decrease or no change in the stock market value depending on the economy news processed.

The time series analysis method, "acceleration" should not be considered because of its unsuitable data output. The acceleration values calculated are either 0 or so close to 0, so the data set expectation was not realistic. This is the reason of high success rate on the acceleartion analysis. On the other hand rest 9 methods are suitable for the correlation and the highest success is achieved from the Bollinger Band with 52% correctly classified news.

The value of success is highly related with the market structure so the success rate here should not be understood as the success rate of the methodology or the classifier. The success rate in the table is the correlation between economy news and the stock market closing values.

## 6. CONCLUSION

During this study, it is first time the effect of time series anlaysis methods over the stock market closing values and their correlation with the economy news in the Turkey case has been studied. The feature extraction method and classification methods are kept simple and the study is mainly focused on the time series analysis. The analysis has shown that the success of Bollinger band is higher than the rest.

We believe this study would help to understand the market strength in Turkey from a financial perspective and also the study can help further research with other classification algorithms and feature extraction methodologies.

## 7. REFERENCES

*AsiaPac Finance (*2013), *List of Technical Analysis Trading Indicators for Stocks and Forex*, http://www.asiapacfinance.com/ trading-strategies/technicalindicators [Accessed February 2013].

Braun, Helmut and John S Chandler (1987),"Predicting stock market behavior through rule induction: an application of the learning-fromexample approach." Decision Sciences 18, No. 3,pp. 415-429.

Fung, Gabriel P. C, Jeffrey X Yu and Wai Lam (2002), "News sensitive stock trend prediction." *Lecture* Notes *in Computer Science* (Springer-Verlag London, UK) ,Vo. 233, pp. 481– 493.

Halgamuge, Saman, Y Zhai and Arthur Hsu (2007), "Combining News and Technical Indicators in Daily Stock Price Trends Prediction." *Advances in Neural Networks - ISNN 2007 (Lecture Notes in Computer Science).* Springer-Verlag Heidelberg, pp. 1087-1096.

Hecht, Nielsen Robert (1987),"Kolmogorov's mapping neural network existence theorem." *IEEE First Annual International Conference on Neural Networks.* San Diego, USA,pp. 11-14.

Karaman, Abdullah S. and Tayfur Altiok (2004), "An experimental study on forecast-ing using TES processes." *WSC'04, Proceedings of the 36th conference on Winter simulation*,pp. 437-442.

Keikha, Mostafa, Mark James Carman and Fabio Crestani (2009), "Blog distillation using random walks." *SIGIR'09, Proceedings of the 32nd international ACM SIGIR conference on Research and devel-opment in information retrieval*,pp. 639-639.

Lu, Hsin-Min, Nina WanHsin Huang, Zhu Zhang and Tsai-Jyh Chen (2009), "Identifying Firm-Specific Risk Statements in News Articles." *PAISI '09 Proceedings of the Pacific Asia Workshop on Intelligence and Security Informatics.* Springer-Verlag Berlin, Heidelberg,pp. 42-53.

Mahajan, Anuj, Lipika Dey and Sk Mirajul Haque (2008), "Mining Financial News for Major Events and Their Impacts on the Market." *Web Intelligence and Intelligent Agent Technology, 2008. WI-IAT '08. IEEE/WIC/ACM International Conference on,*pp. 423-426.

Mahmoud, Safaa, and Moumen T. El-Melegy (2004), "Evaluation of diversity measures for multiple classifier fusion by majority voting." *Electrical, Electronic and Computer Engineering, 2004. ICEEC '04. 2004 International Conference on*,pp. 169- 172.

Masud, Mohammad M, et al. (2011),"Detecting Recurring and Novel Classes in Concept-Drifting Data Streams." *ICDM '11 Proceedings of the 2011 IEEE 11th International Conference on Data Mining.* Washington, DC, USA: IEEE Computer Society, pp. 1176-1181.

Mitchell, Tom M (1997), *Machine learning.* New York: McGraw Hill.

Mittermayer, Marc-André (2004) "Forecasting intraday stock price trends with text mining techniques." *HICSS '04 Proceedings of the Proceedings of the 37th Annual Hawaii International Conference on System Sciences (HICSS'04).* IEEE Computer Society Washington, DC, USA, pp. 64-73.

Mittermayer, Marc-André and Gerhard F Knolmayer (2006), "NewsCATS: A News Categorization and Trading System." *ICDM '06: Proceedings of the Sixth International Conference on Data Mining.* IEEE Computer Society, pp. 1002-

1007.

Nikfarjam, Azadeh, Ehsan Emadzadeh and Saravanan Muthaiyah (2010), "Text mining approaches for stock market prediction." *Computer and Automation Engineering (ICCAE), 2010 The 2nd International Conference on,* pp. 256- 260.

*Ocak, Ibrahim and Seker, Sadi Evren (2012),* "Estimation of Elastic Modulus of Intact Rocks by Artificial Neural Network", Rock Mechanics and Rock Engineering (RMRE), Vol. 45, issue 6, pp. 1047-1054

Ocak, Ibrahim and Seker, Sadi Evren (2013), "Calculation of Surface Settlements caused by EPBM tunneling using artificial neural network, SVM and Gaussian Processes", Vol 70, issue 3, pp. 1263-1276

Rachlin, Gil, Mark Last, Dima Alberg and Abraham Kandel (2007), "ADMIRAL: A Data Mining Based Financial Trading System." *Computational Intelligence and Data Mining, 2007. CIDM 2007. IEEE Symposium on.* doi: 10.1109/CIDM.2007.368947, pp. 720 - 725.

Rosenblatt, Frank (1962), "Principles of neurodynamics: perceptrons and the theory of brain mechanisms." Washington: Spartan Books.

Schalkoff, R J (1997), *Artificial neural network.* New York: McGraw Hill.

Schumaker, Robert P. and Hsinchun Chen (2009), "Textual analysis of stock market prediction using breaking financial news:The AZFin Text system." *ACM Transactions on Information Systems (TOIS)* (ACM New York, NY, USA ) ,Vol. 27, No. 2,pp. 1-19.

Seker, Sadi Evren and Diri, Banu (2010) "TimeML and Turkish Temporal Logic", International Conference on Artificial Intelligence (ICAI10), vol. 10, pp. 881-887

Sadi Evren SEKER, Cihan Mert, Khaled Al-Naami, Ugur Ayan, Nuri Ozalp, "Ensemble Classification over Stock Market Time Series and Economy News", IEEE Conference on Intelligence and Security Informatics, (ISI2013), pp. 272-273

Soni, Ankit, Van Eck, Nees Jan and Kaymak Uzay (2007), "Prediction of stock price movements based on concept map information." *IEEE Symposium on Computational Intelligence in Multicriteria Decision Making.* Honolulu, HI, pp. 205- 211.

Tan, Fook Hwa (n.d.), "Interpreting News Flashes for Automatic Stock Price Movement Prediction." Erasmus University Rotterdam.

Trappe, Wade and Lawrence C. Washington (2006), *Introduction to Cryptography with Coding Theory.* Pearson Prentice Hall.

Wasserman, Philip D. and Tom Schwartz (1988), "Neural networks II. What are they and why is everybody so interested in them now?" *IEEE Expert* 3, No. 1,pp. 10- 15.

Wilder, J. Welles (1978), *New Concepts in Technical Trading Systems.*

Wuthrich, B, V Cho, S Leung, D Permunetilleke, K Sankaran and J Zhang (1998), "Daily stock market forecast from textual web data." *SMC98 Conference Proceedings 1998 IEEE International Conference on Systems Man and Cybernetics Cat No98CH36218.* Ieee, pp. 2720-2725.

Yahia, Moawia Elfaki and B. A Ibrahim (2003), "K-nearest neighbor and C4.5 algorithms as data mining methods: advantages and difficulties." *Computer Systems and Applications, 2003. Book of Abstracts. ACS/IEEE International Conference on.* Tunis, Tunisia, pp. 103.